\documentclass[journal=nalefd,manuscript=article]{achemso}

\usepackage[version=3]{mhchem} 
\usepackage{subcaption}
\usepackage{lettrine}

\author{Dinghe Dai}
\author{Richard Ciesielski}
\author{Arne Hoehl}
\author{Bernd Kästner}
\email{bernd.kaestner@ptb.de}
\phone{+49 30 3481 7104}
\fax{+49 30 3481 69 7104}
\affiliation[Unknown University]
{Physikalisch-Technische Bundesanstalt, Abbestr. 2-12, 10587 Berlin}
\author{Dario Siebenkotten}

\affiliation[Unknown University]
{Physikalisch-Technische Bundesanstalt, Abbestr. 2-12, 10587 Berlin}

\title[Core-Shell Nanoparticle Resonances in Near-Field Microscopy]
{Core-Shell Nanoparticle Resonances in Near-Field
Microscopy Revealed by Fourier-demodulated Full-wave
Simulations}

\begin{document}

\begin{abstract}
 
  We present a detailed investigation of the near-field optical response of core-shell nanoparticles using Fourier-demodulated full-wave simulations, revealing significant modifications to established contrast mechanisms in scattering-type scanning near-field optical microscopy (s-SNOM). Our work examines the complex interplay of geometrical and optical resonances within core-shell structures. Using a finite element method (FEM) simulation closely aligned with the actual s-SNOM measurement processes, we capture the specific near-field responses in these nanostructures. Our findings show that core-shell nanoparticles exhibit unexpected distinct resonance shifts and massively enhanced scattering driven by both core and shell properties. This investigation not only advances the understanding of near-field interactions in complex nanosystems but also provides a refined theoretical framework to accurately predict the optical signatures of nanostructures with internal heterogeneity.
\end{abstract}

\lettrine[lines=1,lhang=0.1]{E}{ngineered} nanoparticles are garnering increasing interest in the fields of medicine~\cite{167,168}, biosensing~\cite{169,170}, catalysis~\cite{171,172}, energy storage~\cite{173}, and opto-electronics~\cite{174} due to their unique properties that arise from their small size and large surface area. The usefulness and functionality of engineered nanoparticles are primarily influenced by the chemical characteristics of their surfaces.
For instance, in medical diagnosis and treatment, core-shell nanoparticles are of special interest as the surface of simple nanoparticles (the core) can be functionalized (the shell) to bind to drugs and deliver them in a targeted manner~\cite{NanoparticlesInMedicineReview, NPFunctionalization}.
Therefore, accurately quantifying both the functionalization and geometry of nanoparticles is crucial for ensuring their optimal performance. 
To achieve this, techniques such as X-ray Photoelectron Spectroscopy (XPS)~\cite{Belsey2016} and Quantitative Nuclear Magnetic Resonance (qNMR)~\cite{Kunc2021} are currently being advanced to meet stringent metrological standards. Optical methods have also been considered as they are non-destructive and chemically specific~\cite{mcquillan2001probing, rotzinger2004structure, hug2006infrared, young2009adsorption, Mudunkotuwa2014}. However, their resolution is typically limited by diffraction to about half the wavelength of the light used, restricting their ability to characterize individual nanoparticles. On the other hand, scattering-type scanning near-field optical microscopy (s-SNOM) provides sub-diffraction spatial resolution and is not limited by the wavelength used. Moreover, this technique simultaneously captures the particle's geometry, allowing the study of correlations, e.g., between particle size and functional group concentration.

\begin{figure}
    \centering
       \includegraphics[width=0.45\linewidth]{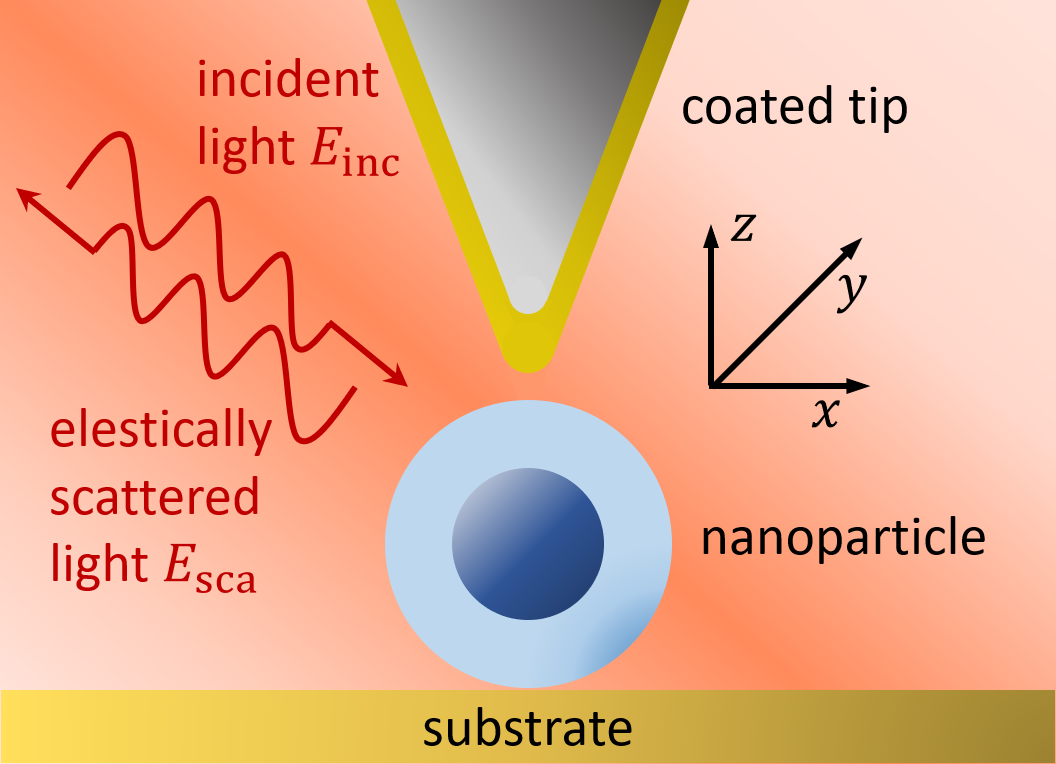}
\caption{
Schematic diagram of the s-SNOM measurement principle applied to a core-shell nanoparticle. The focused radiation, depicted with a red color gradient, displays a relatively broad intensity distribution. A metal-coated tip functions as an optical antenna, directing the radiation into the nanoparticle positioned beneath it. The light scattered from this interaction is analyzed spectrally to reveal both the material composition and geometric properties of the nanoparticle.
}
    \label{fig:intro}
\end{figure}

In principle, s-SNOM~\cite{109} promises access to the degree of surface functionalization through the use of tightly confined optical near-fields~\cite{AshNicholls}, in particular in the mid-infrared spectral range. The confinement in s-SNOM is achieved by focusing electromagnetic radiation onto the metalized probe-tip of an atomic force microscope (AFM) positioned near the sample, which is sketched in Fig.~\ref{fig:intro}. The light is reflected back to the probe in dependence of the sample's geometrical and optical properties, and the scattering from the probe-tip is measured. The sensitivity of s-SNOM on the optical properties of nanoparticles has been shown~\cite{GarciaEtxarri,Stanciu2020,Brehm2006NPs} down to nanoparticles of a few nanometers in diameter~\cite{Cvitkovic_MatSpecNPs,103,Frommelius_AntimonyNPs,Jung_NPs}, including in a spectroscopic manner~\cite{104,Nishida2024_Proteins}. However, comparatively little work has been published on core-shell nanoparticles in s-SNOM to date \cite{110, 130}. This is partly due to the complexity of quantitative descriptions of the corresponding intricate interplay between geometrical and optical factors. Only a few approaches for the modelling of the s-SNOM response of nanoparticles~\cite{EFDM,Nishida2024_Proteins} and core-shell nanoparticles~\cite{Datz_GeneralizedMie} have been published, all approximating the tip with a comparatively small conducting spheroid~\cite{026}. An alternative to the (semi-)analytical s-SNOM modelling approaches - which require at least partial, highly challenging redevelopment when adapted to new geometries, and make approximations about the tip shape - is the use of numerical simulations~\cite{032, 033, 028, 037,120, 035,112,Peng2023, 031, 123}.

To suppress the far-field background dominating the scattered near-field signals, experiments use periodic tip oscillations and evaluate the higher harmonics of the scattered fields~\cite{PDM,AshNicholls}. This approach needs to be mimicked in the simulations. Mooshammer~\emph{et al.}~\cite{037} suggested the application of this procedure to each point of the simulated field, from which insight into the origin of the scattering behaviour of the sample structure, such as a core-shell nanoparticle, can be gained. Extending this approach, we develop a finite element method (FEM) simulation procedure for cylindrically symmetric samples that is closely orientated on the real s-SNOM measurement process, verify it on nanospectroscopy data and apply it to explore the rich interplay of material and geometrical resonances of core-shell nanoparticles on a substrate with varied material and geometrical properties. \\[1cm]

One approach that has been widely used for fast simulations is the employment of simple tip geometries, such as spheres or ellipsoids, in conjunction with non-planar tip geometries~\cite{032, 033, 028}. Conversely, other approaches simulate the entirety of a realistic tip using either the Finite Element Method (FEM) on planar \cite{037,120, 035} or single-step samples~\cite{112,Peng2023} or the simpler, but computationally more efficient, Method of Moments \cite{031, 123} on planar samples. 

We choose FEM as the method of choice as it is known for its accurate determination of the electric field even inside complex structures  \cite{166} and combine it with a realistic tip setup and core-shell nanoparticles on top of a substrate. The scattered field phasor to be determined can be expressed as the sum of the demodulation orders of background- and near-fields at frequencies \(n\Omega\)
\begin{equation}
     E_\mathrm{sca} = \sum_n \tilde{E}_{n}  \exp(i n\Omega t) = \sum_n (\tilde{E}_{\mathrm{nf}, n} + \tilde{E}_{\mathrm{bg},n}) \exp(i n\Omega t), 
    \label{eq:ssnomesca}
\end{equation}
with the tip oscillation frequency $\Omega$ and where \( \tilde{E}_{\mathrm{nf}, n} \) dominates over \( \tilde{E}_{\mathrm{bg},n} \) at higher demodulation orders $n$, due to the rapid decay of the near-fields \cite{PDM,AshNicholls}.

By using interferometric detection schemes such as nanoFTIR \cite{huth2012} and pseudoheterodyne detection \cite{OcelicPSHet}, one obtains a field-sensitive detector signal, $V_D = \kappa E_\mathrm{sca}$, with $\kappa$ being a typically unknown proportionality constant. As $\kappa$ is hard to determine experimentally, the measurement is usually related to a known reference material measured under the same conditions via 
\begin{equation}
    \frac{S_n}{S_{n,\mathrm{ref}}} \exp{(i(\phi_n - \phi_{n, \mathrm{ref}}))}  =  \frac{\tilde{E}_n}{\tilde{E}_{n, \mathrm{ref}}},
    \label{eq:measurand}
\end{equation}
with the demodulated detector voltage $\tilde{V}_n = \kappa \tilde{E}_n \equiv S_n \exp({i\phi_n})$. The left-hand side represents the experimentally measured near-field contrast, while the right-hand side can be predicted by theoretical calculations.

For these theoretical calcuations we employ the commercial FEM solver JCMsuite \cite{jcm} to solve Maxwell's Equations in the frequency domain.  We model the full tip as a 20-µm long cone with a rounded edge (radius = 1000 nm) and a rounded tip apex (radius = 25 nm). The cone features an opening angle of 30° and is comprised of a silicon core and a 70 nm thick gold layer coating.  The system is illuminated by an infrared plane wave incident at a 30° angle to the sample surface. More simulation details are shown in Supplemental Information~S1.

\begin{figure}

    \centering
    \begin{subfigure}{0.27\textwidth}
        \caption{}
        \includegraphics[width=\linewidth]{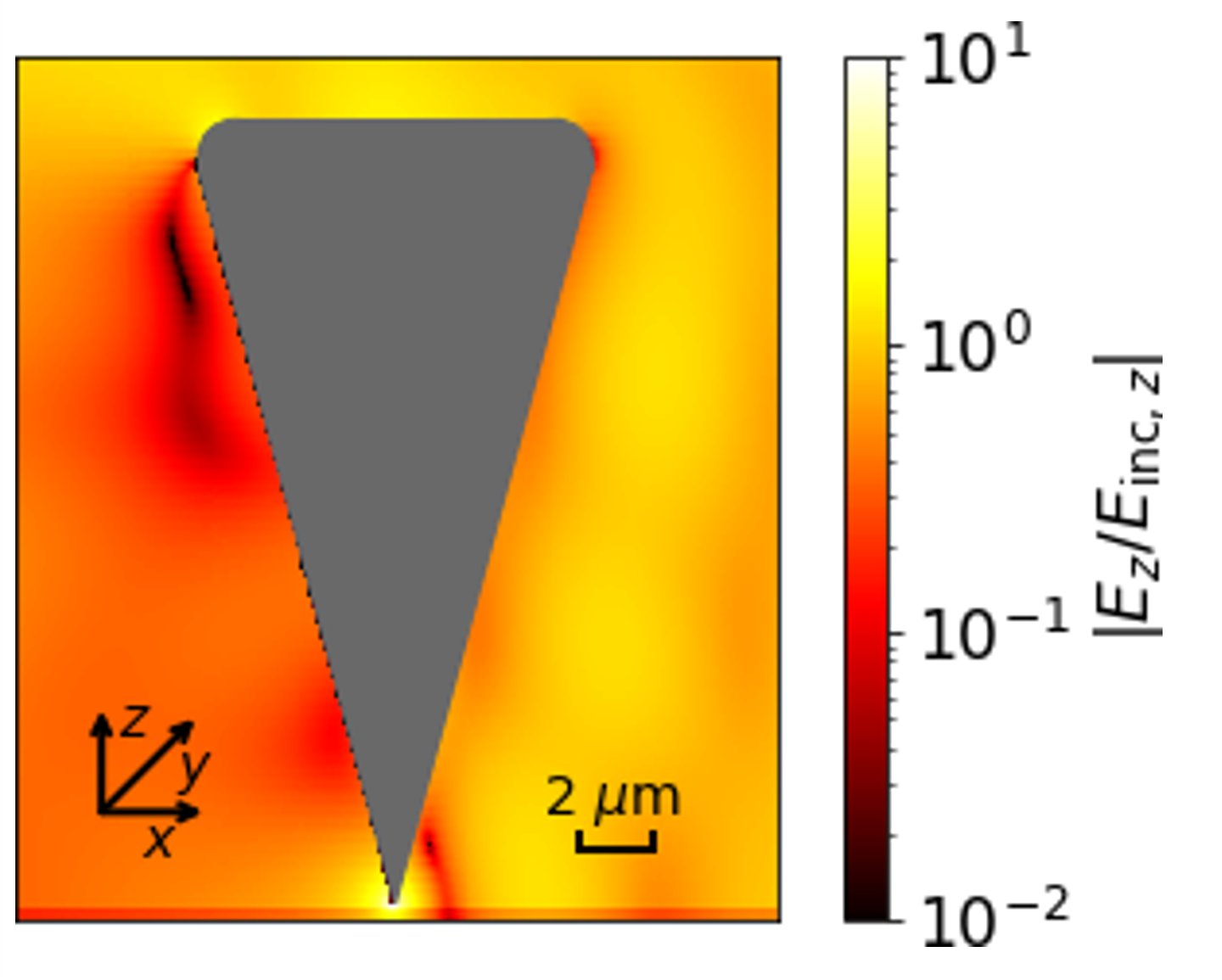}

        \label{fig:undemamp}
    \end{subfigure}
    \hspace{0.03\textwidth}
    \begin{subfigure}{0.27\textwidth}
        \caption{}
        \includegraphics[width=\linewidth]{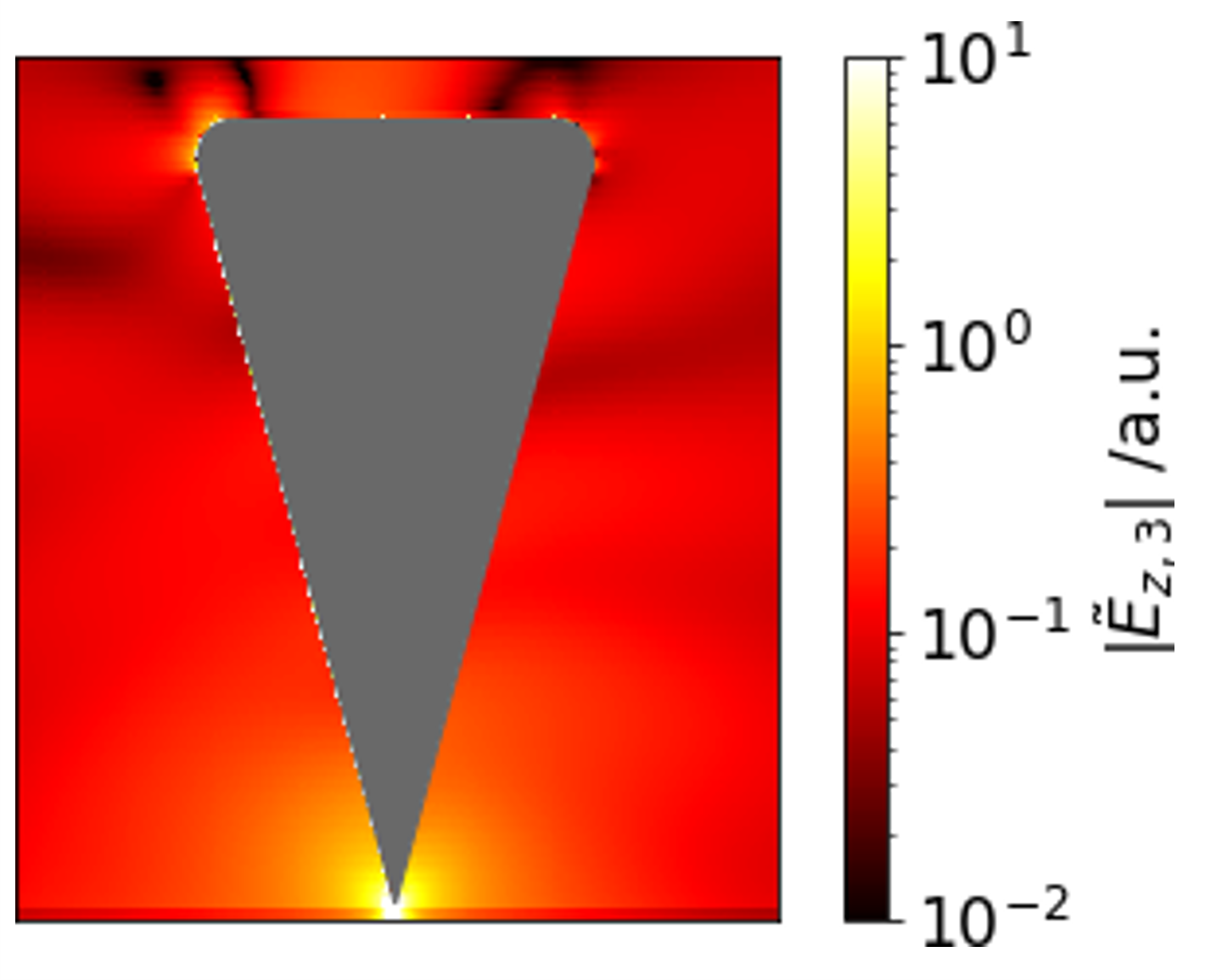}

        \label{fig:demamp}
    \end{subfigure}
    \hspace{0.03\textwidth}
    \begin{subfigure}{0.27\textwidth}
        \caption{}
        \includegraphics[width=\linewidth]{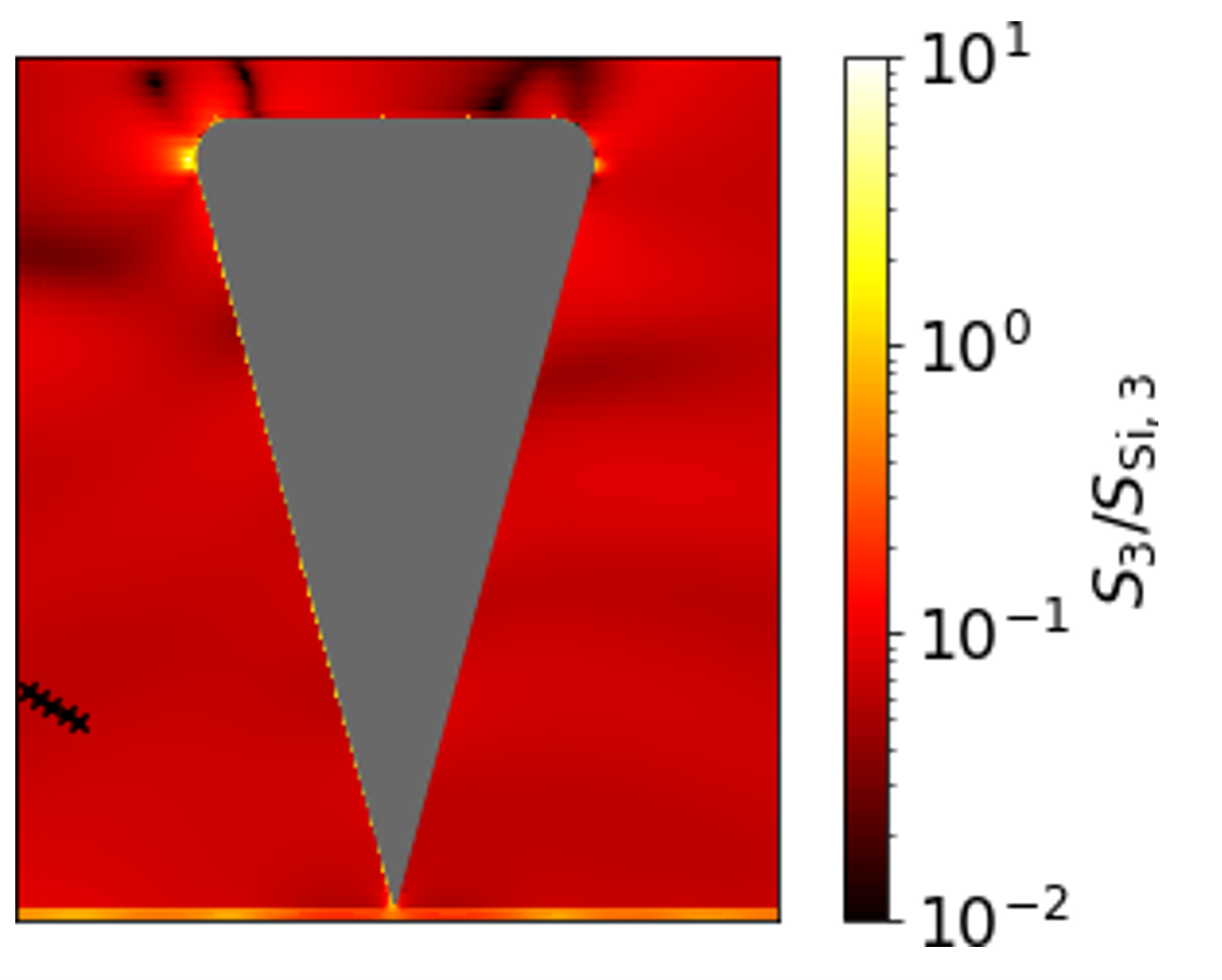}

        \label{fig:reamp}
    \end{subfigure}
    \begin{subfigure}{0.27\textwidth}
        \caption{}
        \includegraphics[width=\linewidth]{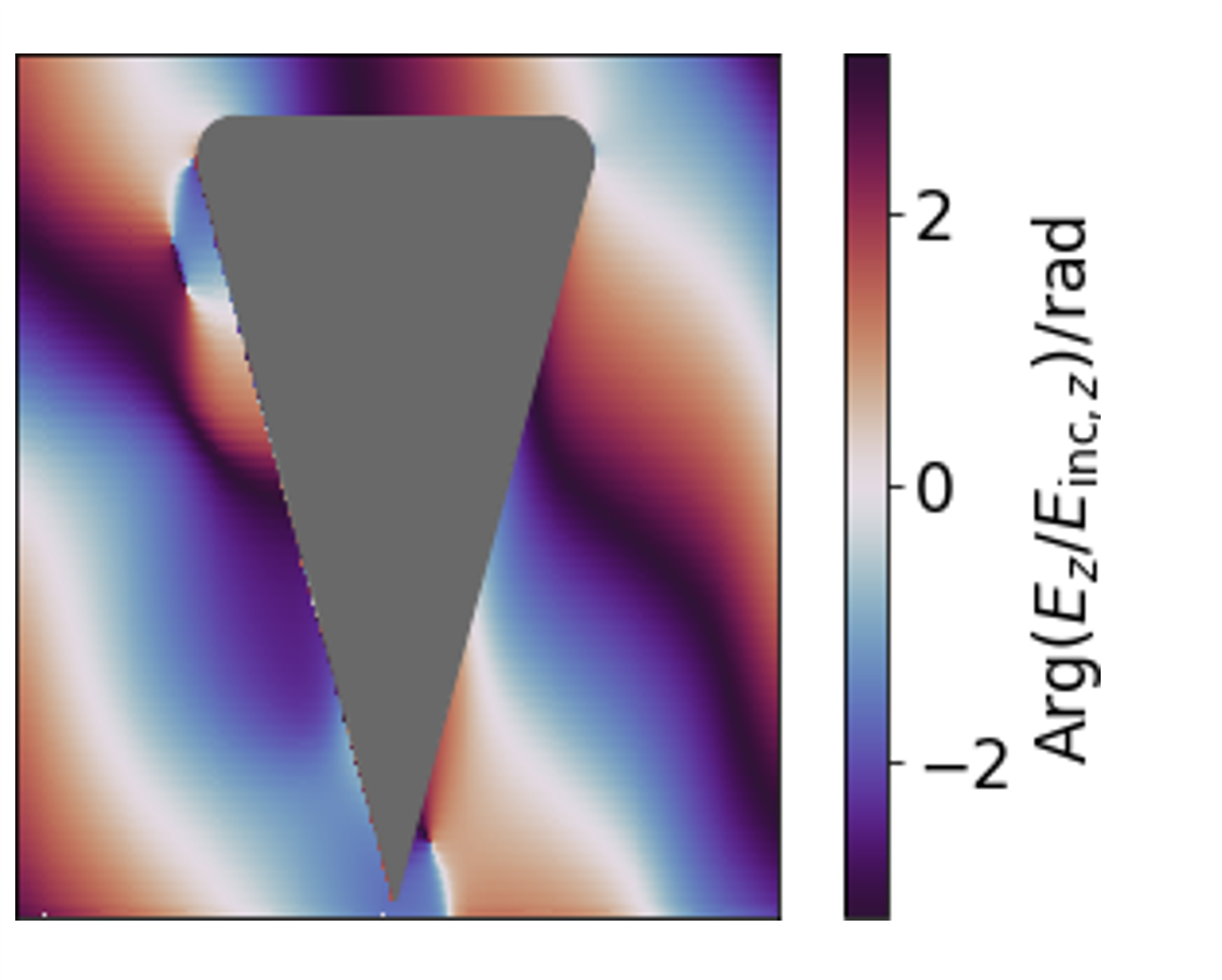}

        \label{fig:undempha}
    \end{subfigure}
    \hspace{0.03\textwidth}
    \begin{subfigure}{0.27\textwidth}
        \caption{}
        \includegraphics[width=\linewidth]{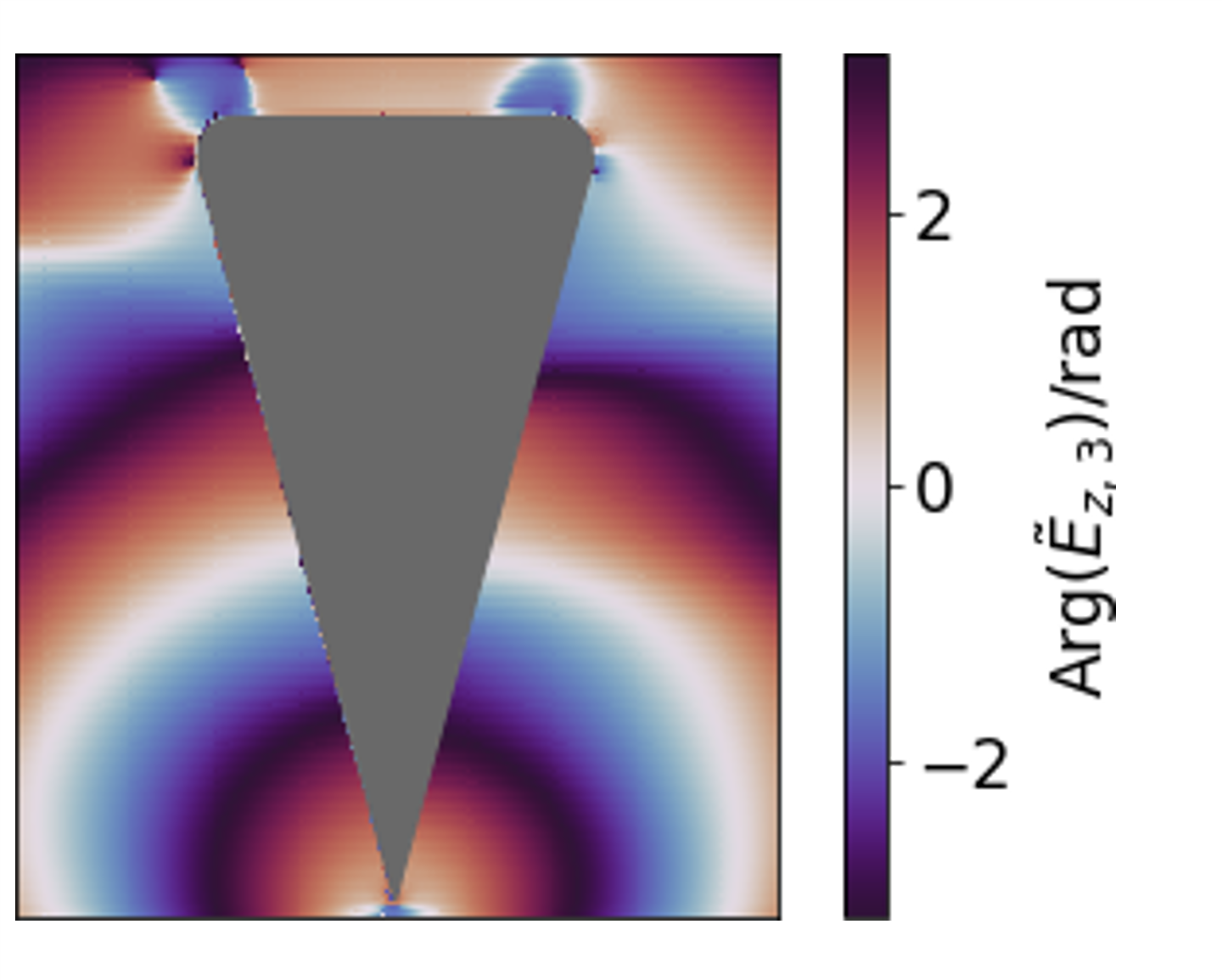}

        \label{fig:dempha}
    \end{subfigure}
   \hspace{0.03\textwidth}
    \begin{subfigure}{0.27\textwidth}
        \caption{}
        \includegraphics[width=\linewidth]{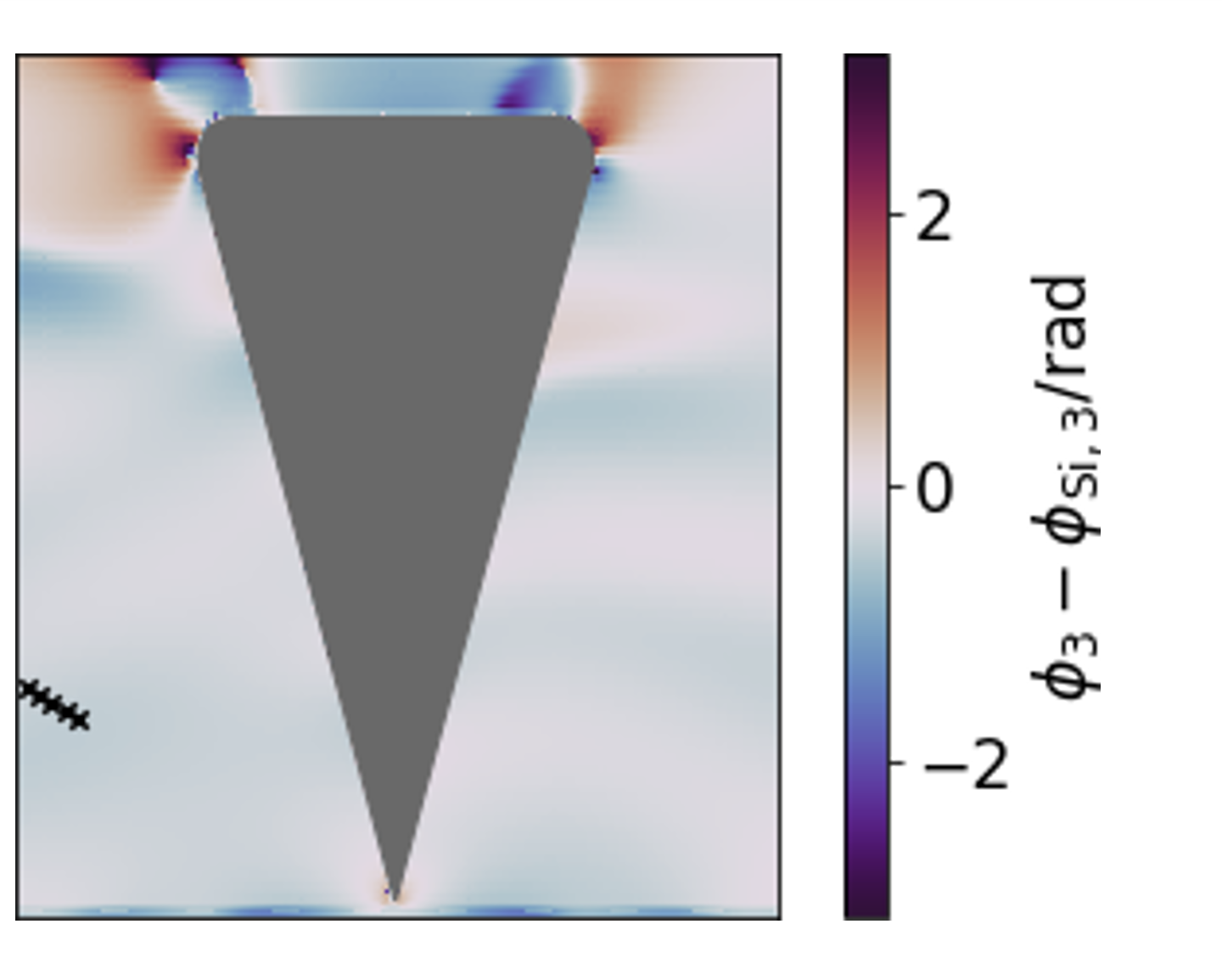}

        \label{fig:repha}
    \end{subfigure}
     \caption{The $z$-component electric field distribution. The amplitude (a) and phase~(d) of the electric field at a probe sample separation of 5 nm. The amplitude (b) and phase (e) of the field Fourier demodulated by varying the tip-sample separation from 2 to 360\,nm. The amplitude (c) and phase (f) of the electric field Fourier demodulated and referenced on silicon. The five black points in demodulated plots (c), and (f) equally distributed in 30° to $x$ axis direction are the simulated detectors where the amplitude and phase signals are captured.}
         \label{fig:demodulation}
\end{figure}

The $z$-component of the electric field (along the long tip axis preferred by the polarization~\cite{sSNOM_p_pol}) of an example of such a simulation (wavelength $\lambda=10\,$µm) is depicted in Fig.~\ref{fig:undemamp} (amplitude) and \ref{fig:undempha} (phase), normalized to $E_\mathrm{inc}$. The tip-sample separation is set to 5$\,$nm and the substrate's optical properties to  $\varepsilon = 1.9 + 0.003i$. Fig.~\ref{fig:undemamp} shows a significant field enhancement between the tip and the sample, while the signals further away are the combined effects of far- and scattered near-fields.

Next, we calculate the Fourier-demodulated field~\cite{037} 
for a sinusoidal tip-sample variation. The total demodulated field can be expressed as:
\begin{equation}
\mathbf{E}(x,z, t) = \sum_n \mathbf{\tilde{E}}_n(x, z) \exp(i n \Omega t).
\end{equation}

To establish Eq.~\ref{eq:measurand} we additionally calculate the Fourier-demodulated field of a reference sample $\mathbf{\tilde{E}}_{n, \mathrm{ref}}(x, z)$ with optical properties defined by its permittivity $\varepsilon_{ref}$. 
The demodulated $z$-component field maps, referenced to silicon ($\varepsilon_{ref}$=12), are shown in Fig.~\ref{fig:reamp} (amplitude) and Fig.~\ref{fig:repha} (phase). They display a relatively homogeneous distribution, as the normalization accounts for the near-field decay. There, the scattered field defined by the near-field contrast can be measured. For robustness, we place five detection points (marked in black on the lower left in Figs.~\ref{fig:reamp} and~\ref{fig:repha}) in the forward scattered direction and calculate the average and standard deviation of the extracted fields. The process of demodulation, normalization and use of detection points away from the tip apex aligns the simulations closely with real-world measurements.\\[1cm]

To demonstrate the methods predictive power for s-SNOM measurements, we compare our method to experimental data obtained from a strongly doped silicon sample, previously presented in~\cite{siebenkotten2024calibration}.  
For the s-SNOM measurements, we used a commercial setup (neaSNOM by attocube systems AG), with a gold-coated Si tip (PPP-NCSTAu by Nanosensors\texttrademark) 
operated with synchrotron radiation in the infrared spectral region~\cite{146}.

\begin{figure}[H]
    \begin{subfigure}{0.45\textwidth}
        \caption{}
        \includegraphics[width=\linewidth]{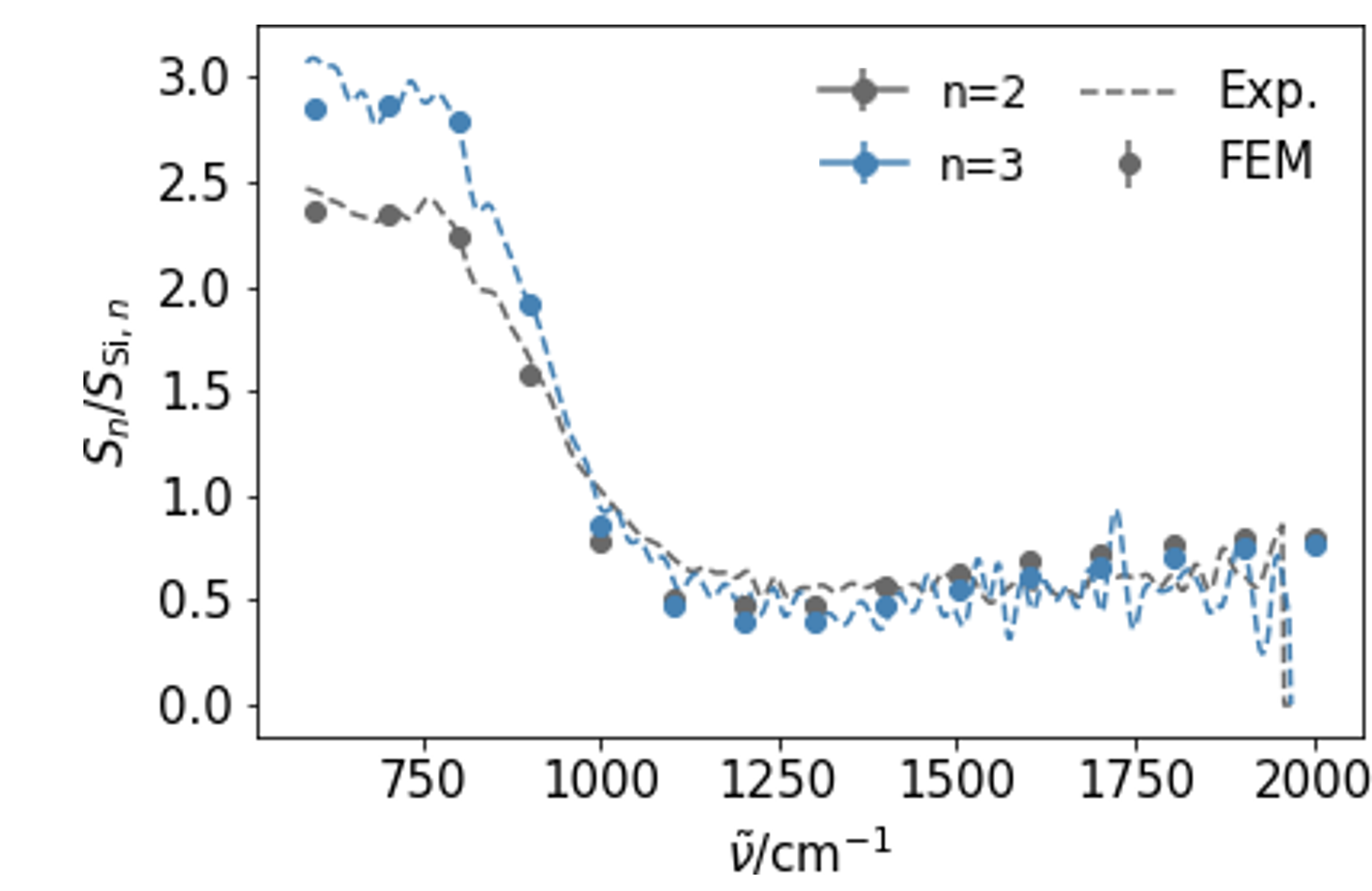}        
        \label{fig:dsiamp}
    \end{subfigure} 
    \hspace{0.035\textwidth}
    \begin{subfigure}{0.45\textwidth}
        \caption{}
        \includegraphics[width=\linewidth]{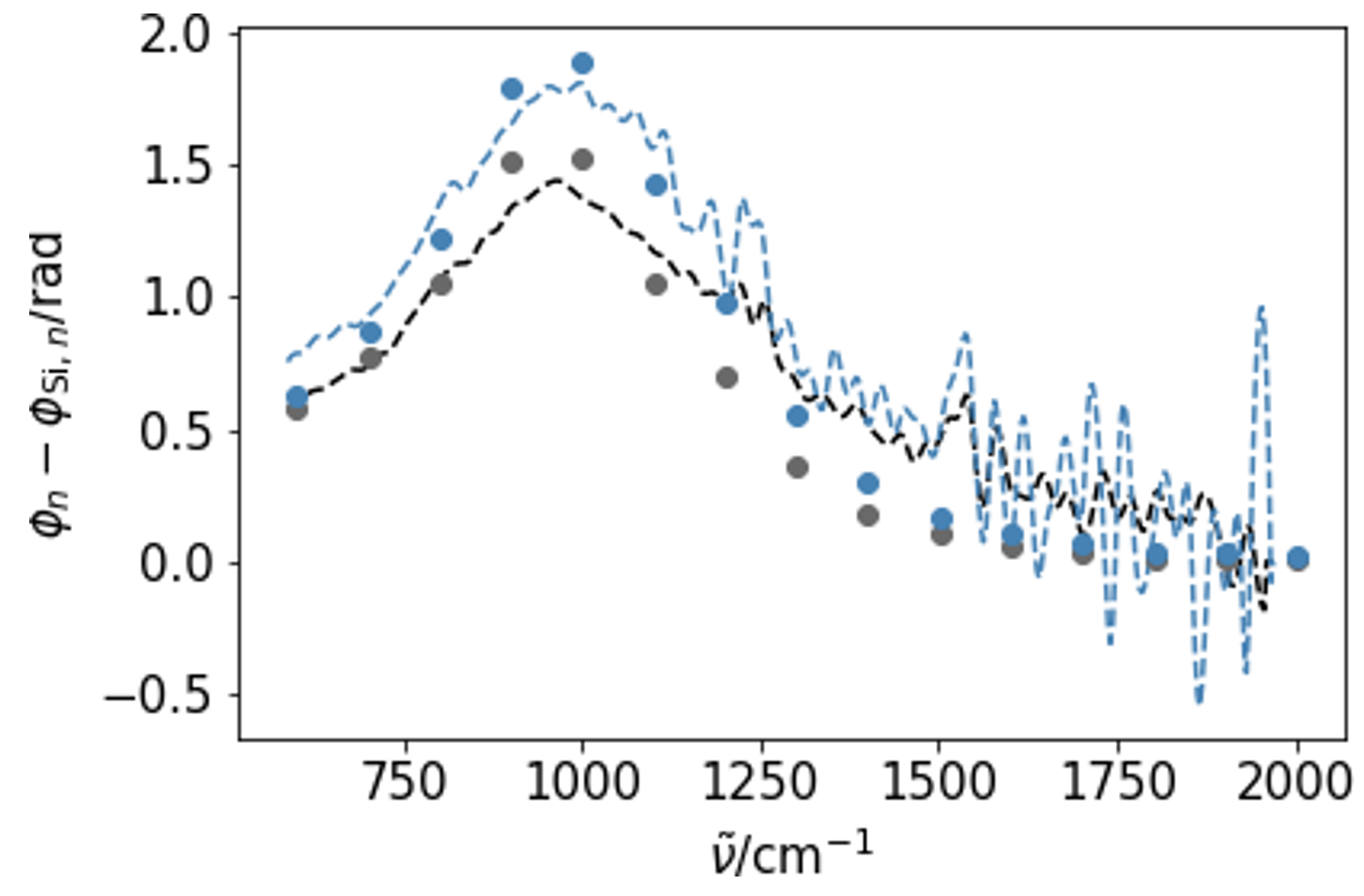}        
        \label{fig:dsipha}
    \end{subfigure}
    \caption{Comparison between the 2$^\mathrm{nd}$ (grey) and 3$^\mathrm{rd}$ (blue) demodulation order amplitude (a) and phase (b) spectra of doped Si measurements (dashed line) and FEM simulations (dots).
     }
    \label{fig:methods}
\end{figure}

Figures \ref{fig:dsiamp} (amplitude) and \ref{fig:dsipha} (phase) depict the measured 2$^\mathrm{nd}$ and 3$^\mathrm{rd}$ demodulation order spectra of the doped Si sample referenced to undoped silicon alongside the theoretical FEM spectra. The calculation assumed the known doping density of $N=4 \times 10^{19}$\,cm$^{-3}$ from which the permittivity has been derived, as described in the Supplemental Information~S2.
Due to the strong doping, the doped silicon acts like a free electron gas with its plasma frequency in the infrared, leading to increased scattering at low wavenumbers. The FEM calculations describe the data excellently, particularly near the plasma resonance at around 950 cm$^{-1}$. Only at higher wavenumbers some systematic offset can be observed, which is small compared to the noise level of the experiment. 
Note that all the parameters used for the modeling of the tip (length, apex radius, opening angle, and coating thickness) were taken from manufacturer specifications and electron microscopy measurements and were not fitted to the data. 
Comparisons to the finite dipole model for several classes of materials are shown in Supplemental Information~S2.

\begin{figure}[htb]
    \centering
    \begin{subfigure}{0.42\textwidth}
        \caption{}
        \includegraphics[width=\linewidth]{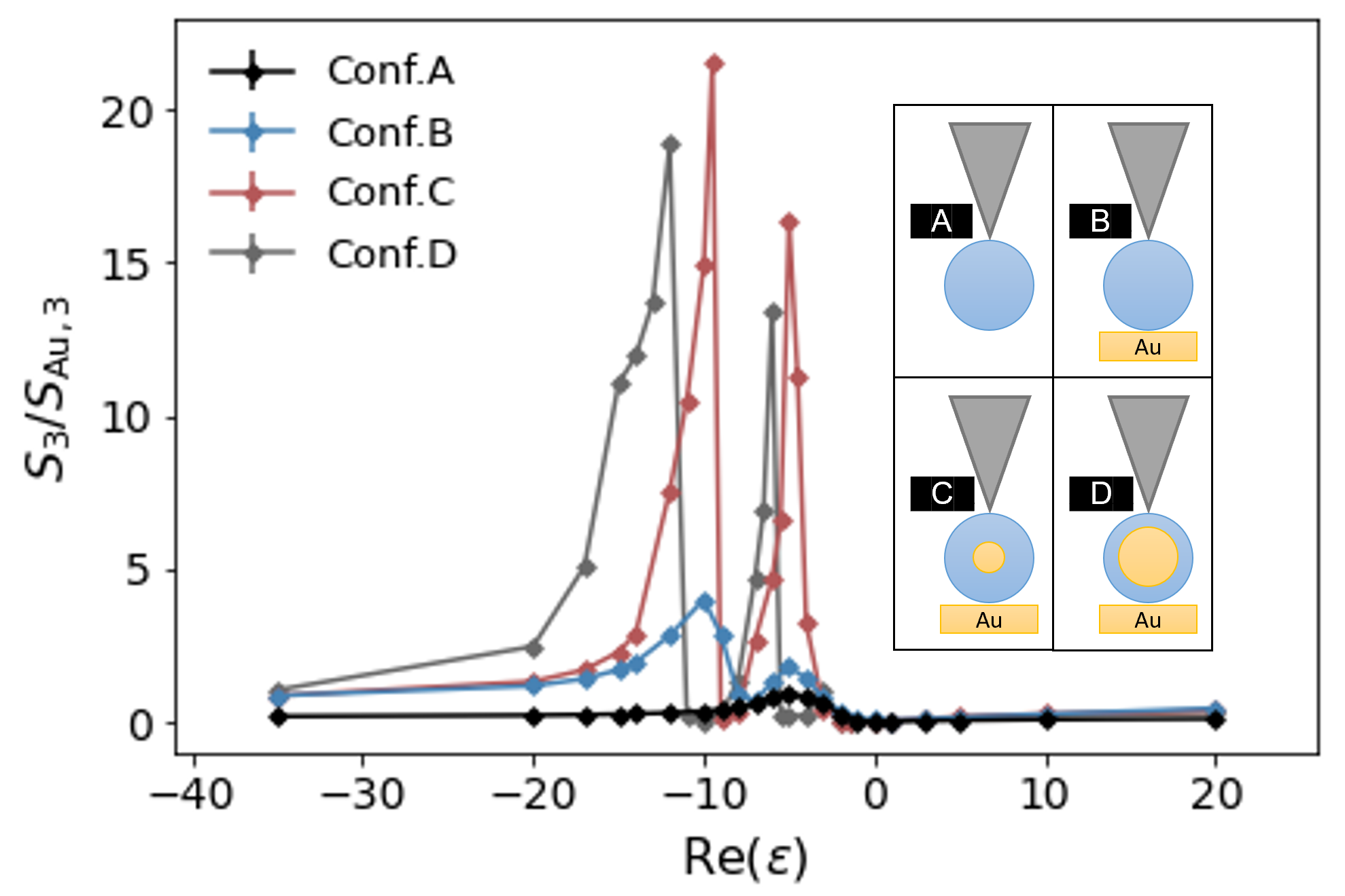}
      
        \label{fig:nanoamp}
    \end{subfigure}
    \begin{subfigure}{0.42\textwidth}
        \caption{}
        \includegraphics[width=\linewidth]{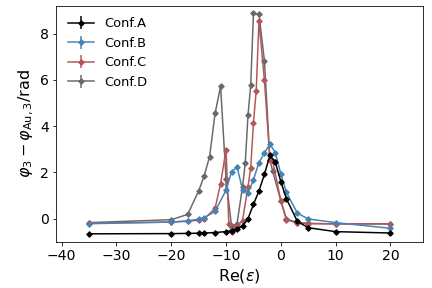}
  
        \label{fig:nanopha}
    \end{subfigure}
    \caption{The amplitude (a) and phase (b) of the third scattering demodulation order obtained from the FEM simulations of 150$\,$nm nanoparticles, for varied real part of the permittivity ranging from -35 to 20. The imaginary part of the permittivity is held constant at 1. Four distinct nanoparticle configurations are represented: the nanoparticle in air (Conf.~A), the nanoparticle on a gold substrate (Conf.~B), the core-shell nanoparticles with 50$\,$nm (Conf.~C) and 100$\,$nm (Conf.~D) diameter gold cores on a gold substrate.}
    \label{fig:nano}
\end{figure}

Next we employ the FEM approach to explore the s-SNOM scattering of single core-shell nanoparticles. 
For clarity we begin with a simpler system consisting of only the tip and a single 150$\,$nm diameter nanoparticle in air (Conf.~A, cf. inset in Fig.~\ref{fig:nanoamp}). The calculated contrast to bulk gold substrate without nanoparticles at 9$\,$µm wavelength is shown in black in Fig.~\ref{fig:nanoamp} (amplitude) and Fig.~\ref{fig:nanopha} (phase) with the real part of the nanoparticle's permittivity $\mathrm{Re}(\varepsilon)$ varied from -35 to 20 and with a constant imaginary part  $\mathrm{Im}(\varepsilon)=1$. Both, the amplitude and the phase, exhibit a single resonance at small negative $\mathrm{Re}(\varepsilon)$, similar to the case of an extended strong oscillator sample as derived in Ref.~\cite{Taubner_PDMReContrast}. Our FEM calculation reproduces the well known result that metals should appear brighter than high-refractive index dielectrics while low-refractive index dielectrics appear dark.
Materials with a small negative $\mathrm{Re}(\varepsilon)$ generate the brightest signal.

Upon the introduction of a gold substrate (Conf.~B, blue) a second peak appears at lower negative $\mathrm{Re}(\varepsilon)$, which is shown by the blue curve in Fig.~\ref{fig:nano}. Furthermore, the contrast is enhanced, which is a known effect of metallic mirrors below nanoparticles~\cite{103}. 
The inclusion of a 50$\,$nm diameter gold core (Conf.~C, red) into the nanoparticle leads to a drastic increase of the resonance scattering and a shift of the peak-position to lower $\mathrm{Re}(\varepsilon)$. When the diameter of the gold core is increased to 100$\,$nm (Conf.~D, grey), that shift is exacerbated. Note that the total diameter of the nanoparticle is kept constant for all four configurations.

\begin{figure}[H]
     \centering
     \begin{subfigure}{0.8\textwidth}
         \centering
         \includegraphics[width=\textwidth]{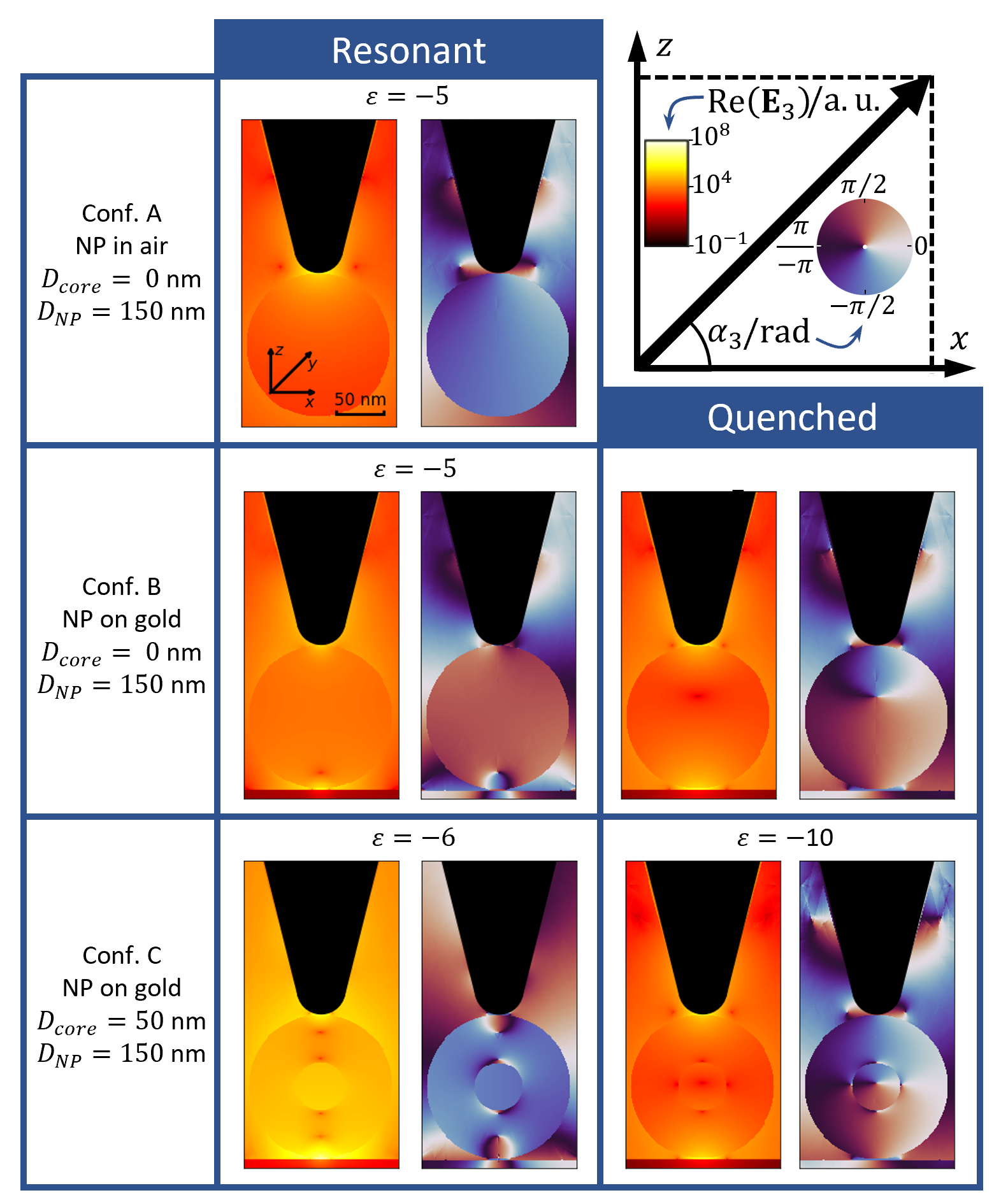}
         
     \end{subfigure}
     \caption{The demodulated electric field distribution at the 3$^\mathrm{rd}$ demodulation order near tip-sample interaction region. Combining the fields in both $x$ and $z$ direction, the field vector Re$(\mathbf{E}_3)$ is investigated in norm and angle (with respect to the $x$-axis). Three 150\,nm nanoparticle configurations are involved: nanoparticles in air (Conf.~A), nanoparticles on a gold substrate (Conf.~B), and core-shell nanoparticles with a 50\,nm gold core on a gold substrate (Conf.~C). For every configuration, the electric field distributions for the resonant (at the peak position near zero in Fig.\ref{fig:nanoamp}) and for the quenched case (at the dip between two peaks) of the permittivity are depicted.}
         \label{fig:results}
\end{figure}

The resonance position shift already shows that both material and geometric properties show non-trivial impact on the scattering. This then poses the question if the quenching between the two peaks is caused by the absence of resonant behaviour, or by an explicit antiresonance. To investigate its origin, Fig.~\ref{fig:results} shows the third demodulation order magnitude ($|\mathrm{Re}(\mathbf{E_3}(x,z)|$) and direction (the angle to the x-axis $\alpha_3$) of the electric vector field in the $x$-$z$-plane for Conf.~A-C for the resonant and quenched case. For Conf.~A, the resonance shows strong enhancement between the tip and the nanoparticle, with a spatially tightly confined rotation of the electric field direction $\alpha_3$. $\alpha_3$ further rotates over larger areas in lobes at the sides of the tip, which are demodulation effects already observed in simulations by Mooshammer \emph{et al.} \cite{037}. When introducing the gold substrate in Conf.~B, a second region of strong field enhancement appears inside the nanoparticle at the substrate-facing side for the resonant case. $\alpha_3$ also fully rotates once throughout that region. In the quenched case, such a region also appears, but flattened between the nanoparticle and substrate. Furthermore, within the nanoparticle the field rotates around a non-central point, with $\alpha_3$ diametrically opposed at opposing sides. For Conf.~C, the fields behave similarly to Conf.~B, but with a centered field rotation. This indicates that the minimum constitutes an antiresonance, as strong field enhancement is present in the near-field, but the opposing directions of the electric field inhibit scattering, leading to the observed minimum at the detection position. These insights highlight the usefulness of the direct access to the electric fields for complex geometries that the FEM calculations provide.\\[1cm]

In conclusion, we analyzed the s-SNOM contrast of core-shell nanoparticles using Fourier-demodulated full-wave simulations. By employing a finite element method (FEM) simulation approach tailored to mimic the real s-SNOM measurement process, we explored the interplay of geometrical and optical resonances within these nanostructures. Our findings reveal that core-shell nanoparticles exhibit resonance shifts and significantly enhanced scattering effects driven by both the core and shell properties, different from the behavior observed in simpler nanoparticle configurations. These results highlight the potential of s-SNOM in the investigation of individual nanoparticles particularly in applications involving functionalized core-shell structures with concurrent correlation to the nanoparticle size. 
While our model as presented is restricted to cylindrically symmetric samples, it can be readily generalized to arbitrary geometries, albeit at increased computational demands. 
This opens the model to more broad applications, such as the analysis of resonance behaviour of complex nanostructures, making it a  useful tool for nanophotonics developments. 
This approach also holds potential for exploring thermal effects in nanostructures, particularly in the context of photocurrent-induced nanoscopy.

\begin{acknowledgement}
This work was supported in part by the Deutsche Forschungsgemeinschaft (DFG, German Research Foundation) through Project-ID452301518 “Investigation of quench switching of antiferromagnets with high spatial and temporal resolution” and Project-ID529998081 “Ultrasensing in the nearfield: polariton enhanced molecular fingerprinting”. SEM Measurements on the AFM tips were performed by Patryk Krzysteczko, which are gratefully acknowledged.

\end{acknowledgement}

\begin{suppinfo}
The supporting information contains additional details of the FEM simulation and a comparison to the finite dipole model (PDF).

\end{suppinfo}

\bibliography{refs}

\end{document}